\documentstyle[11pt,newpasp,twoside,epsf]{article}
\def\edcomment#1{\iffalse\marginpar{\raggedright\sl#1\/}\else\relax\fi}
\reversemarginpar

\begin{document}
\title{OAO1657--415 : A `Missing Link' in High Mass X-ray Binaries?}
 \author{D.H. Maxwell, A.J. Norton}
\affil{Department of Physics and Astronomy, The Open University,
Walton Hall, Milton Keynes MK7 6AA, U.K.}
\author{P. Roche}
\affil{Department of Physics and Astronomy, The University of Leicester,
University Road, Leicester LE1 7RH, U.K.}

\begin{abstract}
OAO1657--415 is only the seventh eclipsing X-ray pulsar known and therefore 
has the potential to yield only the seventh mass of a neutron star in 
an X-ray binary. Here we report photometric and spectroscopic
observations of candidates for the optical counterpart to the system and
identify a B5III or B6V star as a possible companion to the 
neutron star. We
measure the observational parameters of the star and suggest reasons
why OAO1657--415 may be unlike other high mass X-ray binaries. 
\end{abstract}

\section{Introduction}

High mass X-ray binaries (XRBs) consist of a neutron star, usually an 
X-ray pulsar, in orbit around a supergiant or Be-type companion star.
Supergiant XRBs are permanently bright X-ray sources -- they lose mass by a 
roughly spherical wind which may be augmented by Roche lobe overflow.
In contrast, Be star XRBs often exhibit transient X-ray outbursts which
may or may not be correlated with periastron passage through a dense 
circumstellar disc. Using the radial 
velocity amplitude of the companion and the Doppler shift of the X-ray pulsar 
around its orbit, the mass of the neutron star can be unambiguously determined 
if the system is eclipsing.

OAO1657--415 is only the seventh eclipsing X-ray pulsar known and therefore
only the seventh XRB in which the mass of the neutron star may potentially be
measured.  Originally detected by {\em Copernicus} (Polidan et al 1978) it was
at first incorrectly identified with the massive binary V861 Scorpii,
leading to speculation that it
was a black hole system. A more precise source position, provided by {\em HEAO
1} and {\em Einstein}, established that V861 Sco was not the companion and
detected 38.22s pulsations (White \& Pravdo 1979; Parmar et al 1980).  {\em
BATSE} observations of the X-ray pulsar subsequently revealed a 10.4d orbit
with a 1.7d eclipse by the stellar companion (Chakrabarty et al 1993). 
OAO1657--415 falls between recognised classes of object in the Corbet diagram
(Figure 1); and no optical counterpart has been identified.  

\begin{figure}[h]
\plotfiddle{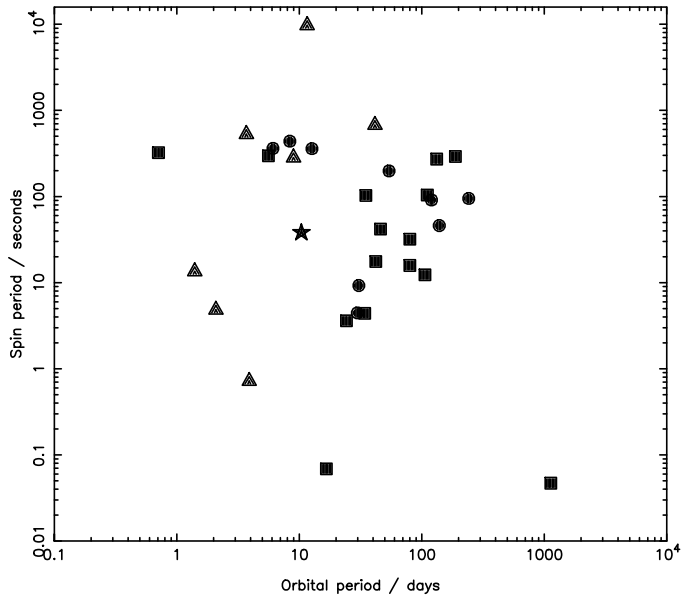}{3in}{0}{85}{85}{-220}{-120}
\caption{The Corbet diagram. Companion stars are shown as supergiants 
(triangles) or Be stars (squares) where spectroscopically identified. Other 
companions are shown as unidentified (circles). Be star XRBs cluster on the 
right of the diagram; supergiant XRBs lie in two regions: wind-fed systems lie 
at the top whilst Roche-lobe fed systems lie towards the left of the diagram. 
OAO1657--415 lies on its own, indicated by a star, in the centre of the 
diagram.}
\end{figure}

\section{Observations}

\begin{figure}
\plottwo{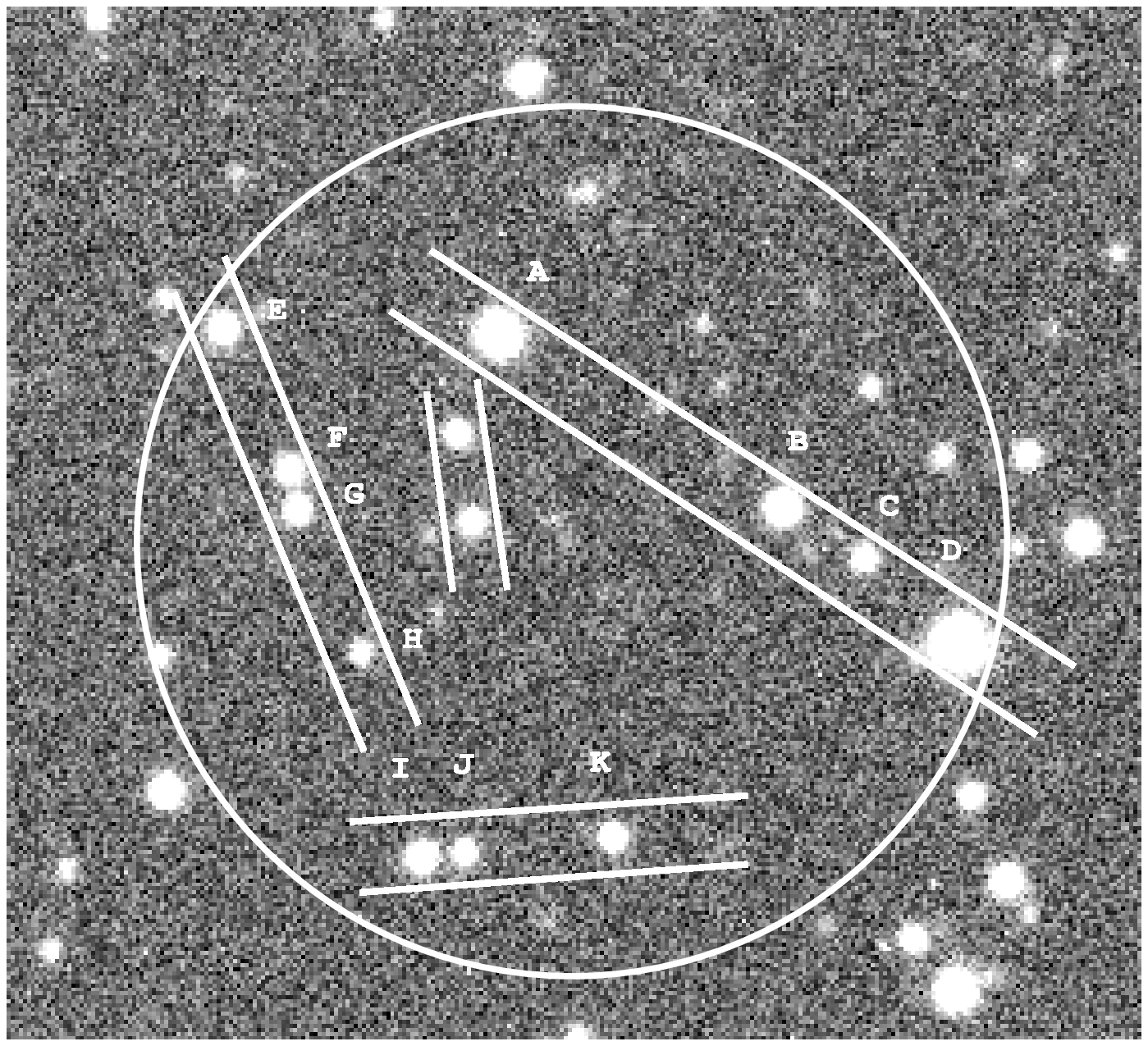}{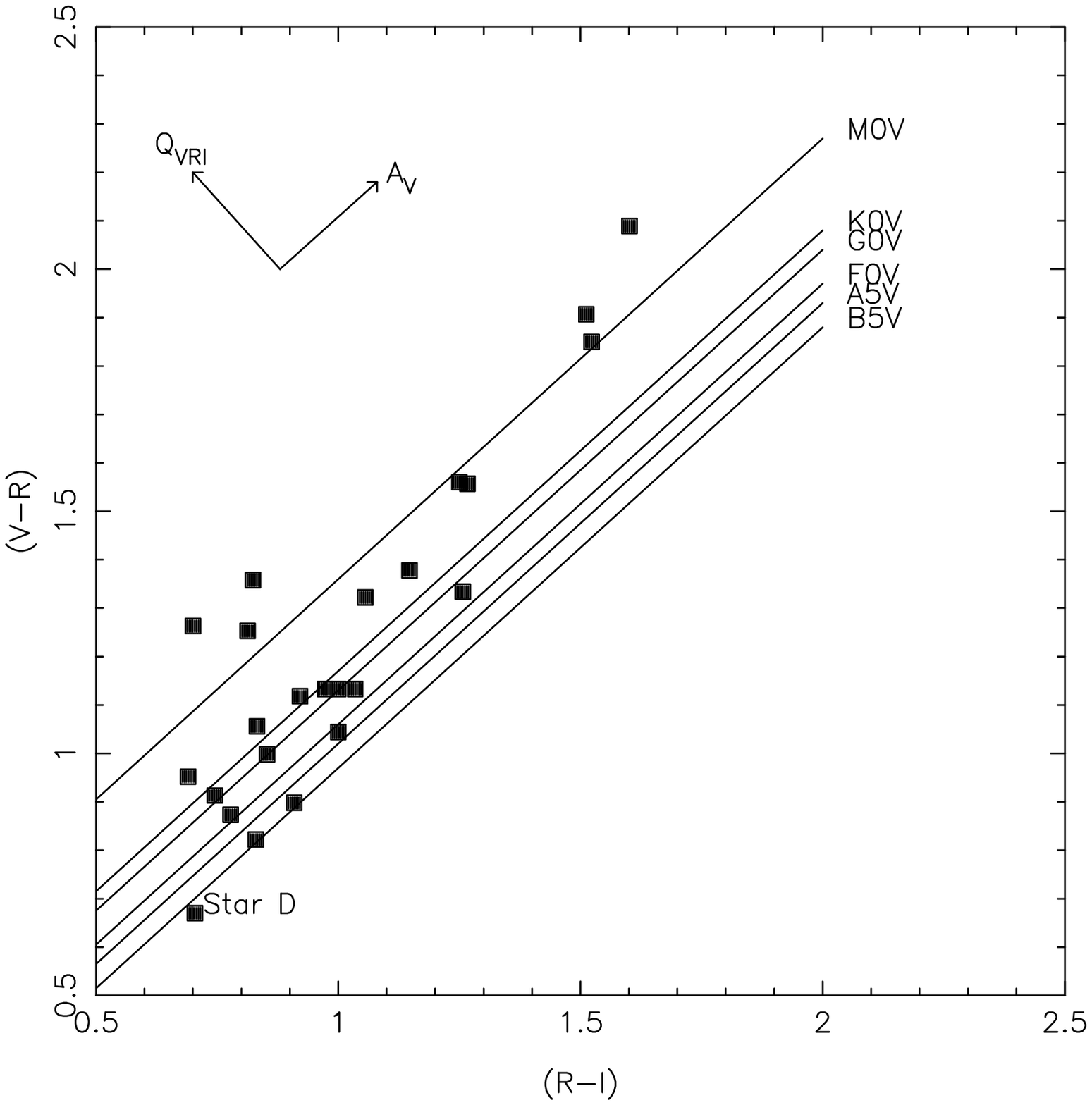}

\caption{(Left) A CTIO V-band image of the field of OAO1657--415 
illustrating the 30$^{\prime \prime}$ {\em Einstein} error circle 
centred at RA: 17\fh 00\fm 47.8\fs, Dec: --41\deg 39\arcmin 14\arcsec (J2000). 
Straight lines show schematic outlines of the slit positions used for the AAT 
spectroscopy reported below. The candidate counterpart is Star D.
(Right) A plot of $(V-R)$ versus $(R-I)$ shows lines of constant $Q\arcmin$
for different spectral types, where $Q\arcmin = (V-R) - 0.91(R-I)$.  Lines 
for luminosity classes III and V are similar.}
\end{figure}

Optical photometry of the X-ray error circle and
surrounding sky (Figure 2a) was obtained by service observations from 
the CTIO 1.5m in the BVRI bands on 12 Aug 1992. 
Infrared photometry was obtained by service observations 
from the AAT 3.9m using IRIS. Aperture photometry was performed on 
all the sources on and within the Einstein error circle, with 
the results shown in Figure 2b.

Spectra were obtained on 12-15 May 1997 using the RGO spectrograph on 
the AAT with the 1200B grating and 25cm camera. 
Spectra cover the wavelength range $\sim 4100$\AA \, to 4900\AA \, at a 
dispersion of 0.75 \AA \, pixel$^{-1}$. Four different slit positions 
(see Figure 2a) were used to obtain spectra of 11 stars within the error 
circle. Only that of Star D (shown in Figure 3) corresponds to an early 
type star. 

\begin{figure}
\plotone{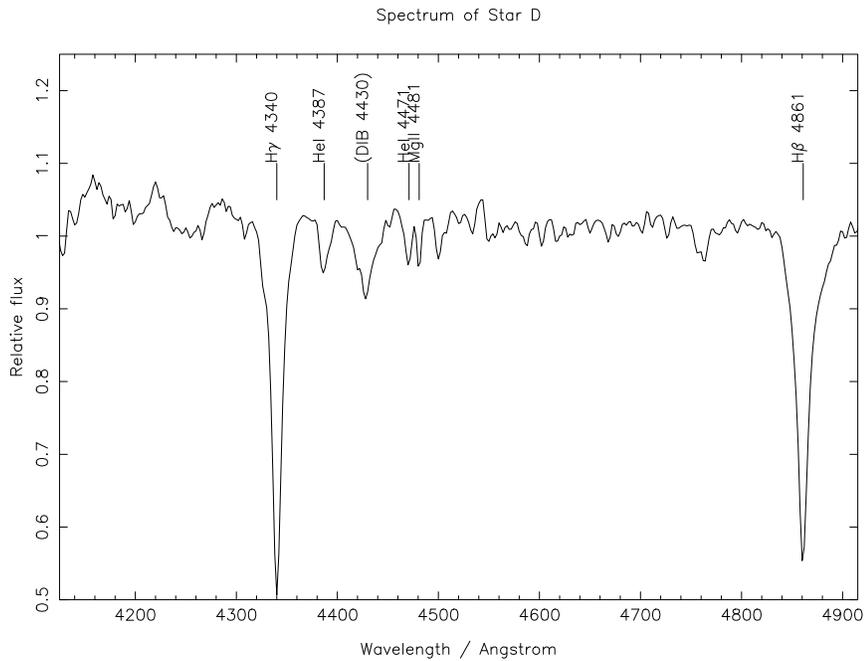}
\caption{The spectrum of Star D}
\end{figure}

\section{Results}

Based on the relative strengths of the He I 4387\AA, He I 
4471\AA \, and Mg II 4481\AA \, lines, the spectrum of Star D 
corresponds to B5III or B6V, the low resolution precludes more
refinement (Cananzi, Augarde \& Lequeux, 1993). Furthermore the 
observed magnitudes of Star D are: U = 16.13 (A. Parmar, private 
communication) B = 15.84, V = 14.84, R = 14.17, I = 13.46, J = 12.59, 
H = 12.23, and K = 12.07. 

The spectral classification is supported by the calculated reddening free 
Q-parameters. Defining $Q = (U-B) - 0.72(B-V)$, we have $Q_{\rm Star D}
= -0.43$, $Q_{\rm B6V} = -0.39$ and  $Q_{\rm B5III} = -0.46$. 
Alternatively, defining $Q\arcmin = (V-R) - 0.91(R-I)$, we have 
$Q\arcmin _{\rm Star D} = +0.03$, $Q\arcmin _{\rm B6V} = +0.05$
and $Q\arcmin _{\rm B5III} = +0.05$.
 
For a B6V/B5III star, M$_{\rm V} \sim -0.90 / -2.40$ (Jaschek \& Gomez 1998) and 
we use the colours of such a star from Wegner (1994) and Johnson (1966).
So for Star D, the colour excess is $E(B-V) = 1.14/1.15$; and the extinction 
is $A_{\rm V} = 3.09 E(B-V) = 3.52/3.55$~magnitudes. 

Using, $m_{\rm V} = 14.84$, $M_{\rm V} = -0.90 / -2.40$ and $A_{\rm V} = 
3.52 / 3.55$ yields a distance for star D of 2.78 kpc assuming it is B6V or 
5.47 kpc assuming it is B5III. Dereddening its colours 
using the extinction law of Wegner (1994) and using these distances, 
we calculate the absolute magnitudes of Star D as shown in 
Figure 4, where they are compared with those of B6V and B5III stars.

\begin{figure}
\plotfiddle{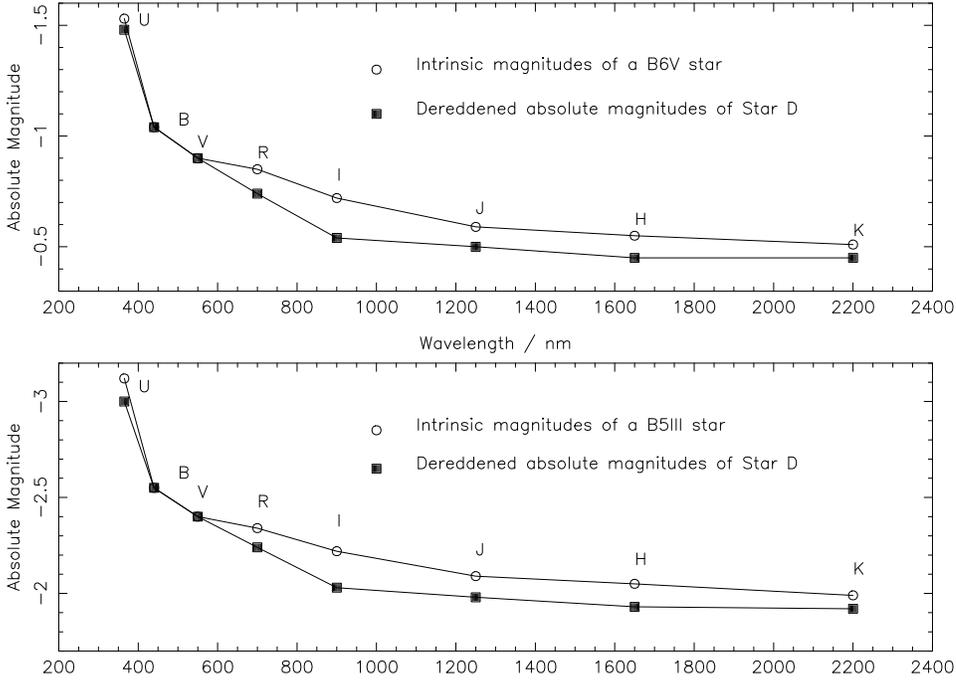}{3.5in}{-90}{50}{50}{-200}{270}
\caption{The de-reddened broad band spectrum of Star D compared with that of
B6V and B5III stars. See text for reddening and distances.}
\end{figure}

\section{Discussion}

Based on the X-ray eclipse properties, Chakrabarty et al (1993) state the 
optical counterpart to OAO1657--415 must have a mass of 14--18M$_{\sun}$ 
and a radius of 25--32R$_{\sun}$, indicating a spectral type of B0-6Ia, and 
implying $M_{\rm V} \sim -6.3$ and $Q_{\rm B0-6I} = -0.92$ to $-0.66$.
No object with these colours is seen in the photometry. 
Assuming an extinction rate of 1.3 mag per kpc (as for Star 
D if it is B6V), a star with $M_{\rm V} = -6.3$ would be visible out to 
$\sim 10$kpc in the V and R bands. Alternatively, 
assuming an extinction rate of 0.65 mag per kpc (as for Star 
D if it is B5III), a star with $M_{\rm V} = -6.3$ would be visible out to 
$\sim 20$kpc in the V and R bands. 
Chakrabarty et al (1993) state that OAO1657--415 must lie at a distance of 
at least 10~kpc, based on the accretion torque during spin up. If this is 
the case it rules out our candidate for the counterpart, and suggests that
the true counterpart must be fainter than $m_{\rm V} \sim 20$. 

Clearly there are problems reconciling our possible counterpart with the
predictions of Chakrabarty et al (1993). Nonetheless, Star D {\em is} the 
most likely counterpart amongst the stars visible in the X-ray error circle
and we reiterate the fact that OAO1657--415 is unusual both in its position 
in the Corbet diagram and in its X-ray properties. The system does not
behave like a typical SGXRB or BeXRB, so we may expect its 
optical counterpart to have unusual parameters too. If Star D {\em is} the 
counterpart, we suggest the following explanation. 

OAO1657--415 is a transient X-ray source and may therefore be a Be star 
system -- the circumstellar material having dissipated at the time of our 
spectroscopy, so yielding a conventional spectrum. Disc loss events
in BeXRBs have been observed on timescales of $\sim 1$~year (e.g. Telting 
et al 1998) so this is feasible. Furthermore, it is possible to reproduce the 
wide eclipse required by the BATSE observations with a disc which is 
optically thick to hard X-rays out to a radius of $\sim 25$R$_{\odot}$ from the 
centre of the star or $\sim 2.5 \times$ the radius of a B5III star.
If the disc in OAO1657--415 had similar unusual density properties to that 
in X~Per (i.e. $\rho_0 = 1.5 \times 10^{-10}$~g~cm$^{-3}$ and $\rho(r) = 
\rho_0 (r/R_*)^{-4.73}$, Telting et al 1998) then $N_H$ is $\sim 
10^{27}$~atoms~cm$^{-2}$ along a line of sight at $\sim 25$R$_{\odot}$ from the 
centre of the star, which is sufficient to absorb hard X-rays. The giant 
classification (rather than supergiant or main sequence Be) may also account 
for the anomalous position of OAO1657--415 in the Corbet diagram. 

As of 2000 October, we have just obtained further spectroscopy of Star D from
the SAAO 1.9m telescope, spread over several weeks during a campaign
to observe SMC X-1. We will use these spectra to search for the expected $\sim
20$~km~s$^{-1}$ radial velocity shifts of the companion star in OAO1657--415. 
If this is confirmed, we will construct a preliminary radial velocity curve of
the companion and so obtain a first estimate of the neutron star mass in
this system.

\end{document}